\documentclass[12pt,fleqn,usenatbib]{mnras}
\usepackage{amssymb,amsmath,graphicx,natbib,longtable}
\usepackage{url,breakurl}
\bibpunct{(}{)}{,}{a}{}{,} 
\usepackage{xcolor,colortbl,tabularx}
\usepackage{booktabs} 
\usepackage{indentfirst} 
\usepackage{braket,newtxtext,newtxmath}

\def \ergsec{\hbox{erg s$^{-1}$}}
\def \ergcms{\hbox{erg cm$^{-2}$ s$^{-1}$}}

\def \chandra{\textit{Chandra}}
\def \xmm{\textit{XMM-Newton}}
\def \sw{{\em Swift}}
\def \fermi{\textit{Fermi}-LAT}

\title[MW observations of the binary MSP PSR J1836-2354A]{Search for multiwavelength emission from the binary millisecond pulsar PSR J1836-2354A in the globular cluster M22}
\author[Amato et al.]{R. Amato$^{1,2.3}$ A. D'A\`i$^{2}$, M. Del Santo$^{2}$, D. de Martino$^{4}$, A. Marino$^{1,2,5}$, T. Di Salvo$^{1}$, \and R. Iaria$^{1}$, T. Mineo$^{2}$
\vspace{6pt}\\
$^{1}$ Universit\`a degli Studi di Palermo, Dipartimento di Fisica e Chimica, via Archirafi 36, I-90123 Palermo, Italy\\
$^{2}$  INAF - IASF Palermo, Via U. La Malfa 153, I-90146 Palermo, Italy\\
$^{3}$ IAAT Universit\"at T\"ubingen, Sand 1, D-72076 T\"ubingen, Germany \\
$^{4}$ INAF - Osservatorio Astronomico di Capodimonte, Salita Moiariello 16, I-80131 Napoli, Italy\\
$^{5}$ IRAP, Universit\`e de Toulouse, CNRS, UPS, CNES, Toulouse, France.\\
}

\begin{document}
\label{firstpage}
\pagerange{\pageref{firstpage}--\pageref{lastpage}}
\maketitle

\date{Accepted ... Received ...; in original form ...}

\pagerange{\pageref{firstpage}--\pageref{lastpage}} \pubyear{}

\begin{abstract}
We present a multi-band search for X-ray, optical and $\gamma$-ray emission
of the radio binary millisecond pulsar 
J1836-2354A, hosted in the globular cluster M22. 
X-ray emission is significantly detected in two \chandra{} observations, performed in 2005 and 2014, at a luminosity of $\sim$\,2--3$\times$10$^{30}$ erg s$^{-1}$, in the 0.5--8 keV energy range. The radio and the X-ray source positions are found consistent within 1$\sigma$ error box. No detection is found in archival \xmm{} and \sw{}/XRT observations, compatible with the \chandra{} flux level. The low statistics prevents us to assess if the X-ray source varied between the two observations. 
The X-ray spectrum is consistent with a power-law of photon index $\sim$1.5. We favour as the most probable origin of the X-ray emission an intrabinary shock scenario. We searched for optical and $\gamma$-ray counterparts to the radio source using data from \textit{Hubble Space Telescope} and \textit{Fermi}-LAT catalogues, respectively. No optical counterpart down to V=25.9 and I=24.7 (3$\sigma$) is detected, which suggests a companion mass of 0.1-0.2 $M_\odot$. Combined with the low X-ray luminosity, this is consistent with a black widow nature of PSR J1636-2354A. Inspecting the 8-year \textit{Fermi}-LAT catalogue, we found a $\gamma$-ray source, 4FGL J1836.8--2354, with a positional uncertainty consistent with the globular cluster, but not with the radio position of the millisecond pulsar.
 \end{abstract}

\begin{keywords}
X-rays: binaries -- X-rays: individual: PSR J1836-2354A -- Stars: pulsars -- Galaxy: globular clusters: individual: M22 (NGC 6656).
\noindent
\end{keywords}

\section{Introduction} \label{introduction}

Millisecond pulsars (MSPs) are neutron stars (NSs) emitting radio pulsed radiation at their spin periods. They can be isolated or in binary systems. According to the recycling scenario \citep{Alpar1982}, MSPs are the outcome of accretion onto the NS of mass transferred from a late-type companion. After Gyr-long mass accretion phase during which these systems appear as low-mass X-ray binaries (LMXBs), the mass transfer rate declines allowing the activation of a radio and/or $\gamma$-ray pulsar powered by rotation of its magnetic field \citep{BhattacharyaVandenHeuvel1991,Burderi2001}. A few systems - three so far - were found to transit from an accretion to a rotation-powered state and viceversa proving the existence of the link between LMXBs and MSPs \citep{Papitto2013Nature,Bassa2014,Stappers2014}.  

Globular clusters (GCs) are the
densest environments in our Galaxy where MSPs can be found. Their high stellar densities imply a high rate of dynamical interactions, such that binary systems are formed through alternative mechanisms to the normal evolutionary channels, e.g. tidal capture \citep{Fabian1975}, collisions with a giant star \citep{Sutantyo1975} or by exchange between primordial binaries \citep{Hills1976}. 
Moreover, due to the aged population, binary systems in GCs are predomintantly constituted of a compact object, like white dwarfs (WDs) or NSs, which accretes matter from its companion, usually a low-mass Main Sequence star. Hence, the X-ray population in GCs is mainly constituted by a mixture of quiescent LMXBs, Cataclysmic Variables (CVs), MSPs and Chromospherical Active Binaries (ABs) \citep[see][for a review]{Heinke2010}.

M22 (NGC 6656) is one of the most luminous GC in the Milky Way.
At a distance of 3.2 kpc, it has a projected core radius ($r_{core}$) of 1.33$^{\prime}$ and 
a half-mass radius of 3.36$^{\prime}$ \citep[2010 edition]{Harris1996}, a tidal radius of 31.9$^{\prime}$ \citep{Alonsogarcia2012}, a total mass of $\sim 5\times10^5\,M_{\odot}$
\citep{Cheng2018} and an absolute age of 12.67 Gyr \citep{forbes10}.
\citet{Lynch2011} reported the detection of two radio MSPs in this GC: J1836-2354A and J1836-2354B. 
J1836-2354A (M22A, hereafter) is a 3.35 ms pulsar
in a binary system with an orbital period of 4.87 h, negligible eccentricity, $a\,sin(i)$=0.046412 lt-s, a mass function of $2.609(1)\times10^{-6}$ and a minimum mass of 0.017 $M_\odot$ for the companion star. An extremely low mass secondary would indicate M22A as a black widow system, rather than a redback system, which instead harbours a non-degenerate secondary (i.e. $M_2\geq0.1M_\odot$) \citep{roberts2018}. The other pulsar (M22B hereafter) is isolated with a 3.23 ms spin period. Both pulsars lie within the cluster core radius.

Besides the radio emission, MSPs can also be detected in other bands,
thus allowing to probe different environments and processes in, or close to, the 
pulsar magnetosphere, e.g. optical emission can come from the companion star or, in the case of a LMXBs, from the accretion disk \citep{Archibald2009}, when present. 

Furthermore, $\gamma$-ray emission from Galactic GCs has been detected by the LAT instrument on board of \textit{Fermi Gamma Ray Space Telescope} (\fermi{}, hereafter) since its launch, in 2008. 
Being MSPs strong emitters of $\gamma$-rays \citep{Chen1991,Harding2005} and being GCs extremely rich of MSPs, the whole $\gamma$-ray emission from GCs is thought to be the convolution of the emission from all the MSPs in a cluster \citep{Abdo2010,Caraveo2014}.
$\gamma$-ray emission from M22 was only recently detected by \fermi{} \citep{zhou2015}, after more than 6 years of observations.  
A flux of $(8.6\pm1.9)\times10^{-12}$ \ergcms{} was derived by fitting the spectrum with a power law model with a spectral index of $2.7\pm0.1$, in the energy range 0.1-100 GeV.

The first X-ray observations of M22 were made with \textit{Einstein} \citep{Hertz1983} and \textit{ROSAT} \citep{Johnston1994}. More recently, \xmm{}  observed the cluster in 2000 \citep{Webb2002,Webb2004} while \chandra{} in 2005 \citep{Webb2013} and in 2014. \cite{Webb2013} analysed the \chandra{} 
observation made in 2005 and
reported a faint X-ray source (Source 3 in their Table 1) as the possible X-ray counterpart of M22A.
We use all the available archival data from \chandra{} and \xmm{}, focusing especially on the longest \chandra{} observation (2014). We also analysed 28 observations performed with the \textit{X-ray Telescope} \citep[XRT,][]{burrows05} instrument on board of the \textit{Neil Gehrels Swift Observatory} \citep[][\sw{} hereafter]{gehrels04}, which has been monitoring the cluster for the past two years.
We also performed a search for the optical counterpart using the \textit{Hubble Space Telescope} (\textit{HST}) catalogue from the HUGS project \citep{Piotto2015}, as well as we inspected the 4-year \fermi{} catalogue (3FGL; \cite{Acero2015} and the 8-yr catalogue (4FGL; \cite{4FGL2019arXiv}).

	\section{X-ray observations and data reduction\label{data}}

\begin{table*}
\centering
\caption{Log of the X-ray observations of M22 analysed in this work.}
\begin{footnotesize}
\begin{tabular}{ccccc}
\toprule
& Obs. & Start Time (UT) & Stop Time (UT) & Exposure Time (s)\\
\midrule
\xmm{} & 0112220201 & 2000-09-19 22:05:00 & 2000-09-20 09:31:56 & 41216\\ 
\midrule
\chandra{} & 5437 & 2005-05-24 21:22:27 & 2005-05-25 02:12:40 & 15819 \\
 & 14609 & 2014-05-22 19:40:24 & 2014-05-23 20:00:44 & 84864\\
\midrule
\textit{\sw{}/XRT} & 34847001&	2017-03-07 06:34:57&	2017-03-07 09:03:36&	2412\\
& 34847002&	2017-03-23 15:07:57&	2017-03-23 19:09:39&	2550\\
& 34847003&	2017-04-03 23:58:57&	2017-04-04 05:22:41&	1988\\
& 34847004&	2017-05-02 03:55:57&	2017-05-02 23:37:16&	2272\\
& 34847005&	2017-05-16 21:24:57&	2017-05-17 00:07:26&	1377\\
& 34847006&	2017-05-30 06:05:57&	2017-05-30 10:24:12&	2926\\
& 34847007&	2017-06-13 19:06:57&	2017-06-13 21:36:51&	3011\\
& 34847008&	2017-06-27 05:14:57&	2017-06-28 00:30:46&	2801\\
& 34847009&	2017-07-11 10:18:57&	2017-07-11 16:41:07&	2821\\
& 34847010&	2017-07-25 01:22:57&	2017-07-26 00:34:23&	2693\\
& 34847011&	2017-08-08 03:15:57&	2017-08-08 16:59:13&	3074\\
& 34847012&	2017-08-22 11:54:57&	2017-08-22 20:47:36&	1529\\
& 34847013&	2017-08-25 11:29:57&	2017-08-25 13:13:34&	925\\
& 34847014&	2017-09-05 13:53:57&	2017-09-05 17:57:26&	1086\\
& 34847015&	2017-09-08 13:16:57&	2017-09-08 15:50:11&	2580\\
& 34847016&	2017-09-19 20:48:57&	2017-09-20 00:23:58&	2580\\
& 34847017&	2017-10-03 03:26:56&	2017-10-03 13:40:01&	2878\\
& 34847018&	2017-10-18 00:14:57&	2017-10-18 23:29:29&	2989\\
& 34847019&	2017-10-31 04:04:57&	2017-10-31 06:33:44&	2221\\
& 10376001&	2018-02-16 02:20:57&	2018-02-17 22:34:10&	8397\\
& 10376002&	2018-03-15 10:04:56&	2018-03-16 00:43:10&	3881\\
& 10376003&	2018-03-16 20:32:57&	2018-03-17 02:13:05&	5305\\
& 10376004&	2018-04-15 02:02:57&	2018-04-15 10:52:39&	5433\\
& 10376005&	2018-04-18 09:53:57&	2018-04-18 13:45:06&	1958\\
& 10376006&  2018-05-15 07:10:57&   2018-05-15 11:24:38&	1645\\
& 10376007&	2018-05-16 07:03:57&	2018-05-17 00:06:40&	7456\\
& 10376008&	2018-06-15 10:51:57&	2018-06-15 19:39:23&	9792\\
& 10376009&	2018-07-15 01:53:56&	2018-07-15 17:07:35&	9816\\
\bottomrule
\end{tabular}
\end{footnotesize}
\label{tab_obs}
\end{table*}

We analysed two \chandra\ observations of M22,
made on 2005 May 24 for 15.82 ks with ACIS-S in the FAINT mode (Observation ID 5437)  
and on 2014 May 22 for 84.86 ks with ACIS-S in the VFAINT mode (ObsID 14609). 
For data extraction and analysis we used CIAO version 4.10 and CALDB version 4.7.7. 
Data sets were reprocessed without including pixel randomization (the parameter \textsc{pix\_adj} was set to \textsc{EDSER}), in order to slightly improve the point-spread function (PSF). 

The \xmm{} observation of M22 was performed on 2000 September 19 (ObsID 0112220201), for a total exposure of 41.2 ks, using the EPIC instruments (pn, MOS1 and MOS2) in imaging mode with the medium filters.
We reprocessed the data to obtain calibrated and concatenated event lists with the Science Analysis Software (\textit{SAS}) version 16.0.0.
We produced images for all the EPIC instruments in three different 
energy ranges: 0.5-2 keV, 2-4 keV, and 4-10 keV.

We analysed all the \sw{}/XRT observations of the source performed between March 2017 and August 2018.
The full XRT observation log 
consists of 28 pointings of 1--3 ks exposure each, with approximately one or two visits per month. All the data were taken in Photon Counting (PC) mode. Data were reprocessed with \textsc{xrtpipeline}  to obtain the cleaned event files and exposure maps, using R.A. and Dec. of the source, as detected in the \chandra{} ObsID 14609 (R.A.\,=\,18:36:25.375, Dec.\,=-\,23:54:51.08, in the J2000 system). We merged all the observations, combined the event lists and exposure maps, using the \textsc{ximage}, version 4.5.1 package. Finally, we extracted the image from the merged event list file.
The log of all the analised X-ray observations is reported in Table \ref{tab_obs}.

\section{Source detection and astrometric corrections of the \chandra{} observation}        
The radio position of M22A determined by \cite{Lynch2011} is 2.2$^\prime$ and 0.9$^\prime$ offset from the \chandra\ pointing directions of the 2004 and 2014 observations, respectively. 
This ensures negligible distortion of the PSF and hence a high accuracy in
determining the position of the source.
For each observation, we created an exposure-corrected 
image and exposure map using the \textsc{fluximage} tool 
with a binning equal to 1; we used the tool \textsc{mkpsfmap} to determine 
the PSF-size at each pixel. We selected two different energy bands, 0.3--10 keV and 0.5--6 keV, and for these bands we set the encircled counts fraction (ECF) equal to 0.5, while the energy of the PSF 
was equal to 1.4 keV and 0.3 keV for the broader and for the softer energy band, respectively.
We used the source detection tool \textsc{wavdetect} with pixel wavelength radii of 1.0, 1.4, 2.0, 2.8, 4.0, 5.6.  The probability threshold was left to the default value of  10$^6$ (corresponding to one spurious source in a 1000$\times$1000 pixel map). Image and detection regions (corresponding to a $3\sigma$ error on the position) are shown in Fig.~\ref{ds9_chandra}. We limited our analysis to the ACIS-S3 chip. 

A X-ray source is found at R.A.\,=\,18:36:25.5(8) and Dec.\,=\,-23:54:51.5(5), with $1\sigma$ errors, in the 2014 observation. The position detected in the 2005 observation differs of 0.1$^{\prime\prime}$ in R.A. with respect to the 2014 one. These are consistent with that reported by \cite{Webb2013}, although with a slightly larger uncertainty, likely due to the different source extraction procedure (ACIS-Extract). The detection is always consistent with a point-like source, with no evidence of extended emission.
The X-ray source is found to be at 0.2$^{\prime\prime}$ East and 0.9$^{\prime\prime}$ North from the radio position of M22A. 
Since the long 84 ks Chandra exposure could be affected by the spacecraft drift, we improved the absolute astrometry,  using a cross-matching method.

For this purpose, we used the UV-optical catalogue of M22 from the \textit{HST} UV Globular Cluster Survey \citep[HUGS;][see Section \ref{optical_analysis}]{Piotto2015,Nardiello2018}, available at the University of Padua\footnote{\url{http: //groups.dfa.unipd.it/ESPG/treasury.php}}.
The catalogue covers an area of about 4$^\prime\times$4$^\prime$, centred on the cluster core. The surveys also encompass two distant regions \citep[parallel fields,][]{Simioni2018}, but none of the X-ray sources detected in the ACIS-S3 chip fall in those two regions. We therefore limited our analysis to the cluster HUGS source catalogue. Among the optical sources, we could select only nine that satisfy the condition of being the only ones bright (typically F814W$<$18 mag) within a small ($\lesssim1.2^{\prime\prime}$ major axis) 1$\sigma$ error ellipse. In most cases, the optical source was the only one (when more than one bright source was present the corresponding X-ray source was disregarded). In just a few cases, two or three much fainter stars were present. The association was done irrespective of being cluster members or not (see also Section \ref{optical_analysis}). Among the nine sources, eight are within the cluster core and one within the half-mass radius. One of them corresponds to the source labelled CV1 by \cite{Webb2013}, classified as a cataclysmic variable through the study of its the X-ray emission and optical spectrum. Its position matches the star R0047833 in the HUGS catalogue. We use the CIAO tools \textsc{wcs\_match}, to perform a  cross-matching through a translation (\textsc{method=trans}), and \textsc{wcs\_update} to upgrade the aspect solution file, the level$=$2 event files and the list of the detected sources. We find an average systematic shift of $+0.071^{\prime\prime}$ in R.A. and of $-0.634^{\prime\prime}$ in Dec., with an rms value of $0.3^{\prime\prime}$. Applying this correction, we then find the X-ray source at R.A.=18:36:25.5 and Dec.=-23:54:52.1. The radio MSP M22A lies well inside the 1$\sigma$ X-ray error ellipse (see Fig.\,\ref{optical_image}). Hence, the detected X-ray source can be confidently seen as the counterpart of the radio MSP M22A. \\

\section{X-ray Data analysis} 
\label{X-ray_analysis}

We find 5.5 and 11.8 net counts for ObsID 5436 and ObsID 14609, respectively. 
The net count rates are then $(4.1\pm1.8)\times10^{-4}$ cts s$^{-1}$ (ObsID 5436) and $(1.8\pm0.4)\times10^{-4}$ cts s$^{-1}$ (ObsID 14609). We verified the consistency of the two count rates by a Poissonian ratio test. We tested the null hypothesis probability of the first rate being equal to the second. The resulting \textit{p-value} of 0.1 does not constitute a strong evidence against the null hypothesis probability, which is not rejected.  
We concluded that there is not any statistically significant variability between the two observations. We also investigated the distribution of the arrival times of the detected photons with energies up to 8 keV, considering an extraction region of 1$^{\prime\prime}$, for both the 2005 and 2014 \chandra{} observations. 
We do not detect any clear modulation linked to the orbital period ($P_b=0.2028278011(3)$ days), possibly due to the very low statistics.\\

We extracted a source spectrum from each observation, selecting a circular area centered at the best-fit position returned by \textsc{wavdetect} using a radius of 1$^{\prime\prime}$ and binning the spectrum to have at least 1 count per noticed bin. 
We used {\sc xspec}, version 12.9, for spectral analysis.
Due to the low number of counts, we used the C-statistic \citep{Cash1979}.
Errors are given at 1$\sigma$ confidence 
level, if not stated otherwise.

Since no statistically significant variability is present in the two observations, we fitted the two spectra together, in the energy range 0.5--6 keV, adopting two alternative models: an absorbed power law and an absorbed black-body. 
We used the \textsc{tbabs} (in {\sc xspec})
component for the interstellar neutral absorption, setting the element abundances from \cite{Wilms} and
 the cross-sections from \cite{Verner}, and  the equivalent hydrogen column density value $N_{\textrm H}$ fixed to $0.197\,\times\,10^{22}$ atoms cm$^{-2}$ \citep{Cheng2018}. 

The power law model gave a photon index $\Gamma=1.5_{-0.6}^{+0.7}$, while the black-body model (\textsc{bbodyrad} in {\sc xspec}) has a best-fitting temperature of $0.8\pm0.4$ keV. 
To evaluate the fit goodness, we iterated over 1000 Monte Carlo simulated spectra, within {\sc xspec}. We obtained the 0.30\% of realisations with lower C-statistic values than the best fit ones, in both cases. Hence, the models are both acceptable, though the very low number of counts does not allow us to discriminate between them.

The unabsorbed fluxes, calculated in the energy range 0.5--8 keV, are $2.3^{+1.2}_{-0.6}\times10^{-15}$ \ergcms{} for the power law model and $1.8^{+1.2}_{-0.9}\times10^{-15}$ for the black-body model.
These values give an X-ray luminosity of 2.8$\times10^{30}$ \ergsec\ for the power law model and 2.2$\times10^{30}$ \ergsec{} for the black-body model, respectively, assuming a distance of 3.2 kpc (see Table \ref{tab_fit_results}).
We obtain a X-ray flux slightly lower than that reported by \cite{Webb2013} of $5.2\times10^{-15}$ \ergcms{} (1$\sigma$ error). This is due to the different power law slope assumed by \cite{Webb2013} in their analysis (2.1 instead of 1.5). 
However, by fitting the 2005 spectrum with a fixed the power law slope at 2.1, we obtained a slightly higher, but still consistent, unabsorbed flux, equal to $9.1\times10^{-15}$ \ergcms{}, in the energy range 0.5-8 keV. \\

\begin{figure*}
\resizebox{\hsize}{!}{\includegraphics[height=.3\textheight]{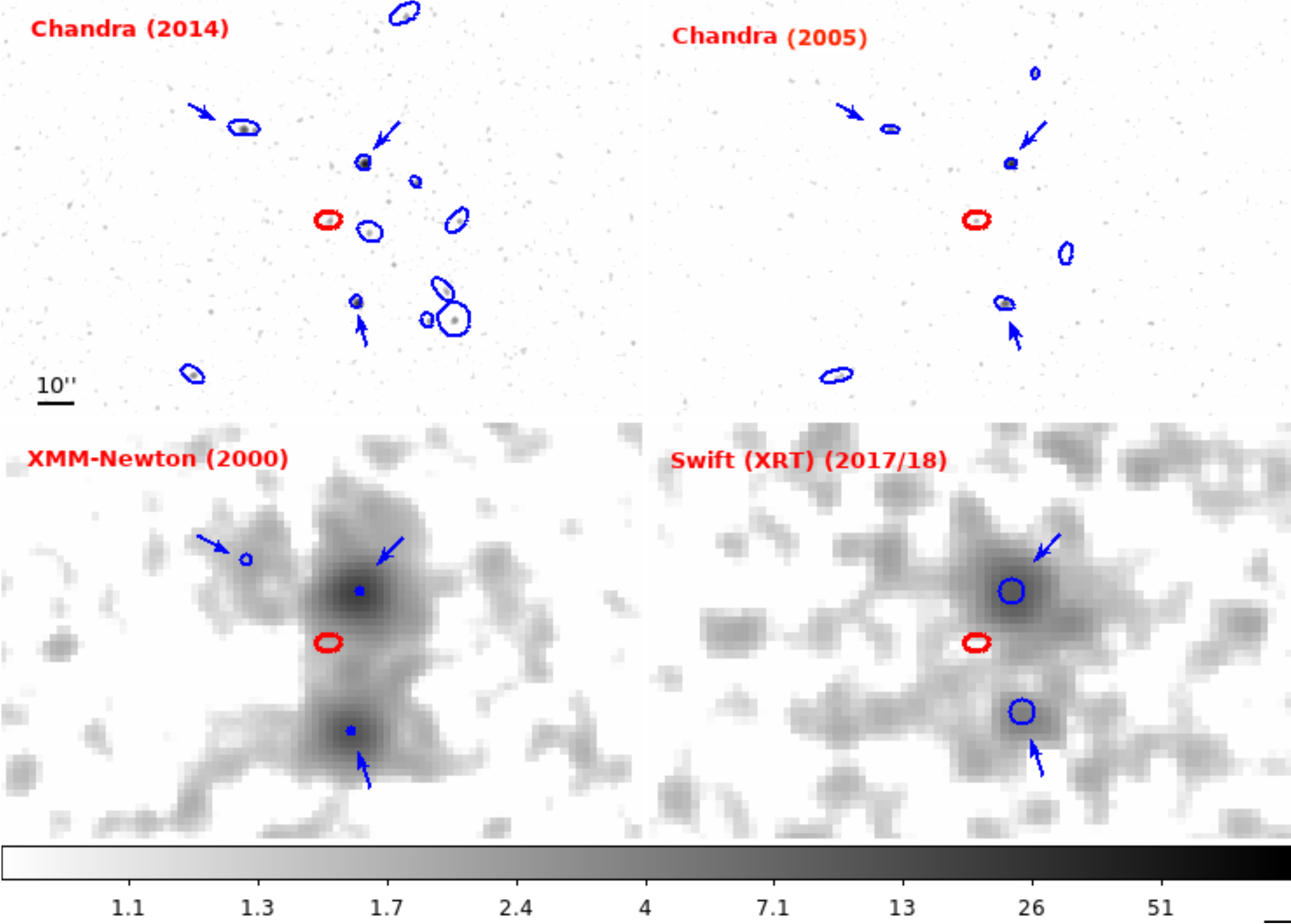}}
\caption{X-ray images of \chandra{} ObsID 14609 (\textit{top left panel}) and 5437 (\textit{top right panel}), of \xmm{} obs. (\textit{bottom left panel}) and of the stacked \sw-XRT observations (\textit{bottom right panel}). The red ellipse corresponds to the position of M22A in the longest \chandra{} obs. (14609), the blue circles/ellipses indicate the other detected X-ray sources. The dimensions of each ellipse in \chandra{} observations correspond to a 3$\sigma$ positional error as given by the detection pipeline, the dimension of the circles of \sw{} observations are given by a centroid procedure and the ones of \xmm{} observations are the catalogued positional errors (\url{http://xmm-catalog.irap.omp.eu/}). The blue arrows point to the most luminous sources close to M22A detected in almost all the data sets.}
\label{ds9_chandra}
\end{figure*}

\begin{table*}
\centering
\caption{Best-fit values of the simultaneous fit of the spectra of M22A from \chandra{} ObsID 5437 and 14609. The fit was performed with the C-statistic, the errors are at 1$\sigma$ confidence level and the goodness was calculated over 1000 Monte Carlo simulations on the ObsID 14609.}
\begin{tabular}{cccccc}
\toprule
Model & $\Gamma$ & $kT$ & $R_{eff}$ 
& Unabs. Flux [0.5-8 keV] & $L_X$ [0.5-8 keV] \\
& & (keV) & ($10^{-3}$ km) 
& ($10^{-15}$ \ergcms{}) & ($10^{30}$ \ergsec)\\ 
\midrule
\textsc{power-law} & $1.5_{-0.6}^{+0.7}$ & & &  $2.3^{+1.2}_{-0.6}$ & $2.8^{+1.5}_{-0.9}$  \\ 
\midrule
\textsc{bbodyrad} && $0.8\pm0.4$ & $6.5_{-3.8}^{+7.5}$ & $1.8_{-0.9}^{+1.2}$ & $2.2_{-1.1}^{+2.0}$ \\
\bottomrule
\end{tabular}
\label{tab_fit_results}
\end{table*}

The archival \xmm{} and \sw{} observations have overall exposure times of $\sim$\,41 ks and $\sim$\,96 ks. Using the NASA's HEASARC tool \textsc{WebbPIMMS}\footnote{\url{https://heasarc.gsfc.nasa.gov/cgi-bin/Tools/w3pimms/w3pimms.pl}}, we estimated the expected count rates for the EPIC instruments and \sw{}/XRT observations. We converted the mean flux of the two \chandra{} observations derived from the power law model into count rates, obtaining $5.4\times10^{-4}$ cts s$^{-1}$ for {\it XMM}/EPICs and $4.4\times10^{-5}$ cts s$^{-1}$ for \sw{}/XRT. The count rate thresholds (3$\sigma$) for \xmm{} observation and for the stacked \sw{} one are of $6.9\times10^{-4}$ cts s$^{-1}$ and $8.5\times10^{-5}$ cts s$^{-1}$. Hence, the source flux is well below the threshold of detectability in both the data sets. Moreover the PSFs are far larger (nominally 15$^{\prime\prime}$ at 1 keV for \xmm{} and 18$^{\prime\prime}$ at 1.5 keV for \sw{},  against $0.5^{\prime\prime}$ of \chandra), so that M22A, which is in the cluster core, cannot be resolved with respect to the closest and brightest source (source 2 of \cite{Webb2013}, see also Fig.~\ref{ds9_chandra}).

However, since it cannot be excluded that the source could have undergone a change of luminosity in the recent past, we inspected the  \sw{}/XRT images one by one, with \textsc{ximage}, using a signal to noise ratio threshold of three. Once we checked out that the source was never detected, we looked for its X-ray emission in the stacked XRT image.
For purpose of comparison with \cite{Webb2004}, we also performed a source detection on the \xmm{}  combined EPIC/pn and EPIC/MOS images, using the tool \textsc{edetect\_chain}, with the appropriate Energy Conversion Factor (ecf) values of the medium filter configuration.
In neither case we detect any source at the radio position of the MSP, as the source have remained below the threshold sensitivity of the two instruments. 
The detection pipelines, indeed, identified sources with fluxes down to $9\times10^{-15}$ \ergcms{} for \xmm{} and to $1.1\times10^{-14}$ \ergcms{} for \sw{}. The sensitivity thresholds, together with the larger PSFs, justify the lack of detection of M22A.
 
\begin{figure}
\resizebox{\hsize}{!}
{\includegraphics[height=.3\textheight,angle=90]{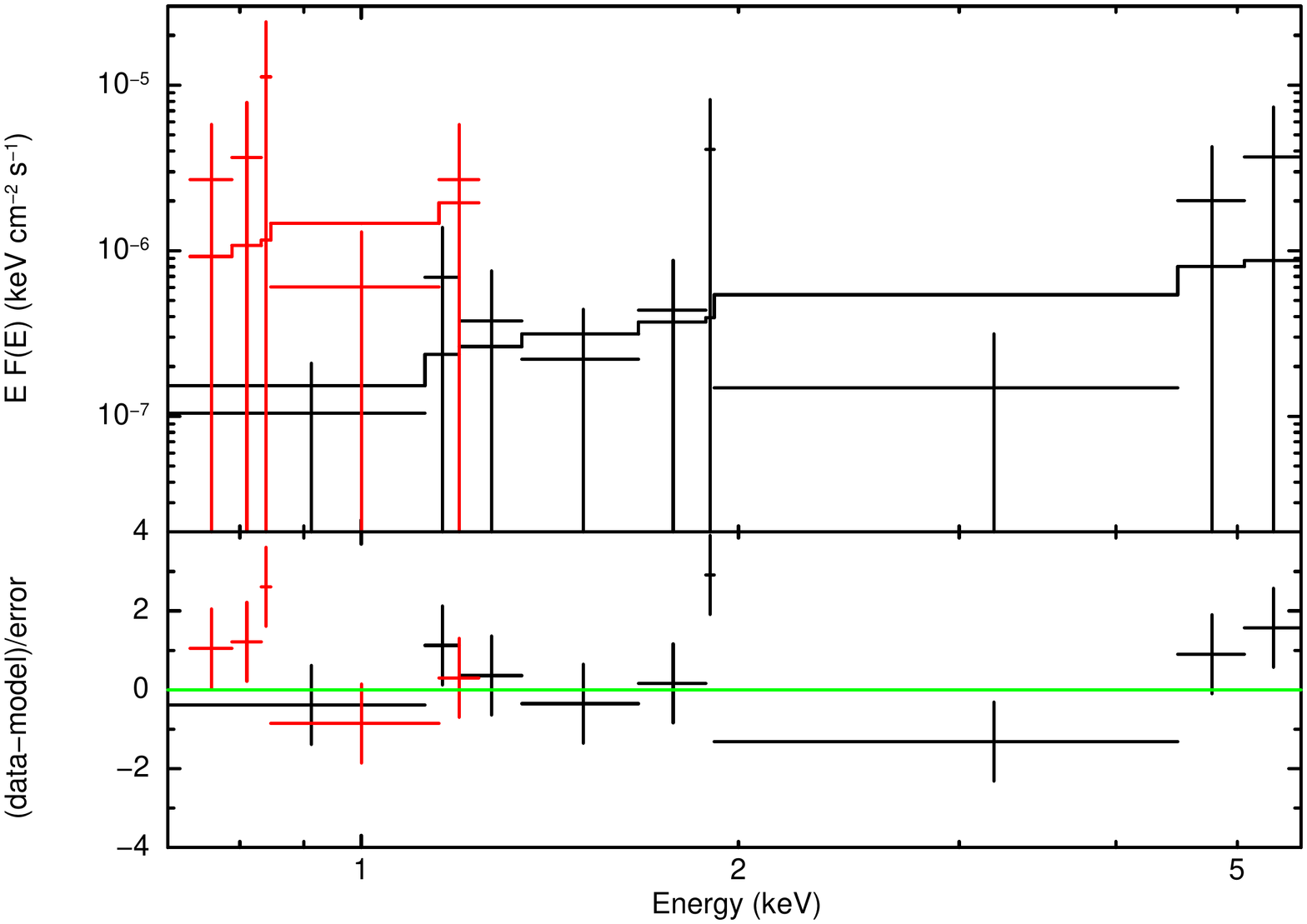}}
\caption{Simultaneous fit of \chandra{} obs. ID 14609 (black) and obs. ID 5437 (red) with a power law plus absorption model and residuals as (data-model)/error where error is calculated as the square root of the model predicted number of counts, in the energy range 0.5-6 keV.}
\label{fit_image}
\end{figure}

\section{Optical observations}
\label{optical_analysis}

\begin{figure*}
\resizebox{\hsize}{!}
{\includegraphics[height=.3\textheight]{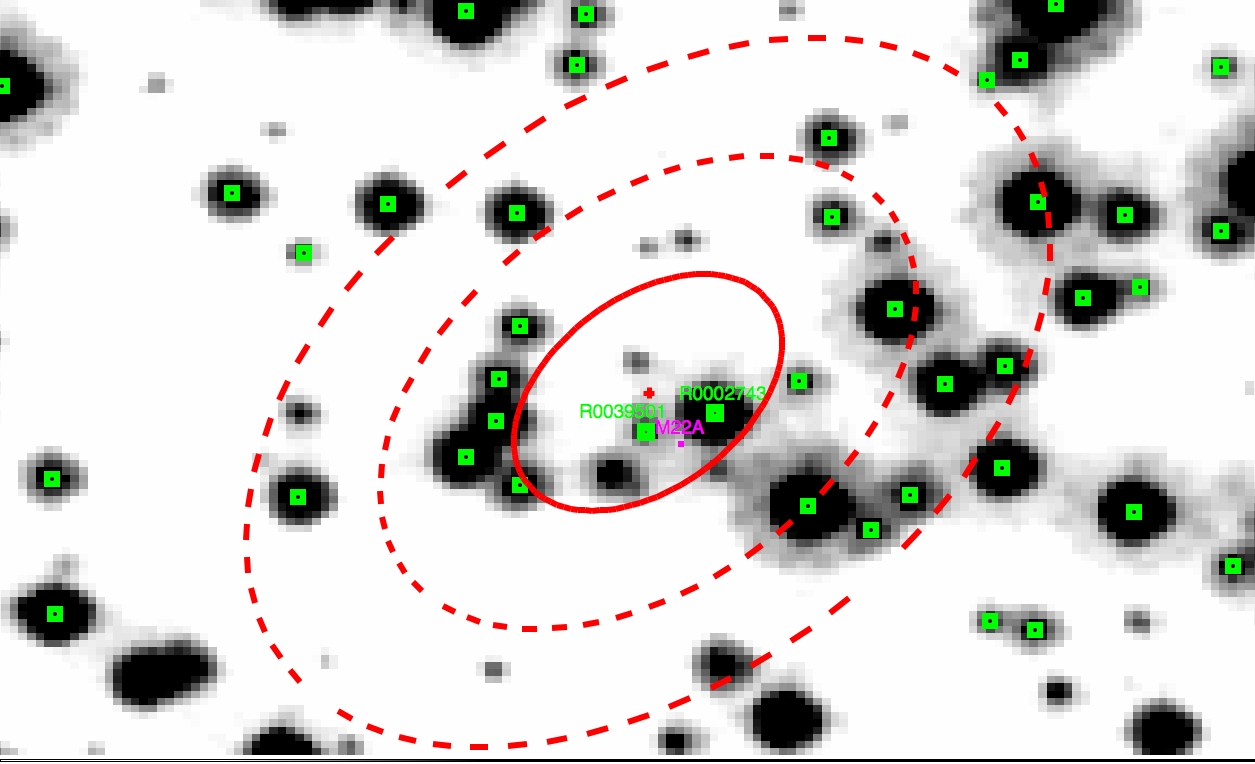}}
\caption{The $8^{\prime\prime}\times4^{\prime\prime}$ enlarged region of the \textit{HST} stacked ACS image in the F818W filter \citep{Nardiello2018} around M22A. North is up, East is left. M22A is marked in magenta. The X-ray 1$\sigma$ error ellipse is reported with a red line, the 2$\sigma$ and 3$\sigma$ error ellipses with red dashed lines. The green boxes mark the optical stars belonging to the M22 cluster with probability membership $>$ 80\%. The two cluster stars, labelled R0039501 and R0002743, have accuracies that rule out any association with M22A.}
\label{optical_image}
\end{figure*}

We searched for the optical counterpart of the radio MSP M22A using \textit{HST} images and the astrophotometric catalogue of M22 \citep{Nardiello2018} from the treasury project HUGS \citep{Piotto2015}. M22 has been imaged in several filters with the WFC3/UVIS (F275W, F336W, F438W) and ACS/WFC cameras (F606W and F814W). We inspected the stacked images in all the five filters, against the astrophotometric catalogue that also provides probability membership for each detected star \citep[see][for details]{Nardiello2018}.
Within the accuracy of the radio position provided by \cite{Lynch2011}, no optical counterpart is detected. The two closest cluster member stars, catalogued as R0039501 ($m_{814w}$ = 20.59(5)) and R0002743 ($m_{814w}$=17.254(7)) in the HUGS project list, are found at much larger distance of 0.197$^{\prime\prime}$ and 0.237$^{\prime\prime}$, respectively. The optical positions of these two stars are very accurate, 0.0014$^{\prime\prime}$ and 0.0024$^{\prime\prime}$ respectively (Nardiello, private communication), and therefore we exclude them as possible counterparts. We infer a 3$\sigma$ upper limit at the position of the radio source of $m_{F606W}\geq25.6$ mag and $m_{F814W}\geq24.7$ mag in the stacked long exposures in these two filters.
The stacked astrometrically corrected image in the F814W filter is shown in Fig.~\ref{optical_image}, together with the radio position of the MSP from \cite{Lynch2011} and 
with the X-ray position of our detection in the latest \chandra{} dataset. 

While we are confident that no optical counterpart is detected for the radio source M22A in the \textit{HST} images, we note that \chandra{} error region in Fig.~\ref{optical_image} shows four or five optical sources within the 1$\sigma$ region and tens of sources at the 3$\sigma$ level. A scrupulous inspection of the closest optical sources in the Colour-Magnitude diagram revealed no bona-fide candidate to a possible red straggler source \citep{Geller2017}, which are sometimes associated to quiescent X-ray binary systems \citep{Shishkovsky2018}. We therefore believe that the source identified in the \chandra{} data is the X-ray counterpart of the radio MSP M22A and consequently none of the optical sources in its error ellipse can be safely associated to the X-ray source.

\section{The $\gamma$-ray emission from M22}

\begin{figure}
\resizebox{\hsize}{!}
{\includegraphics[height=.3\textheight]{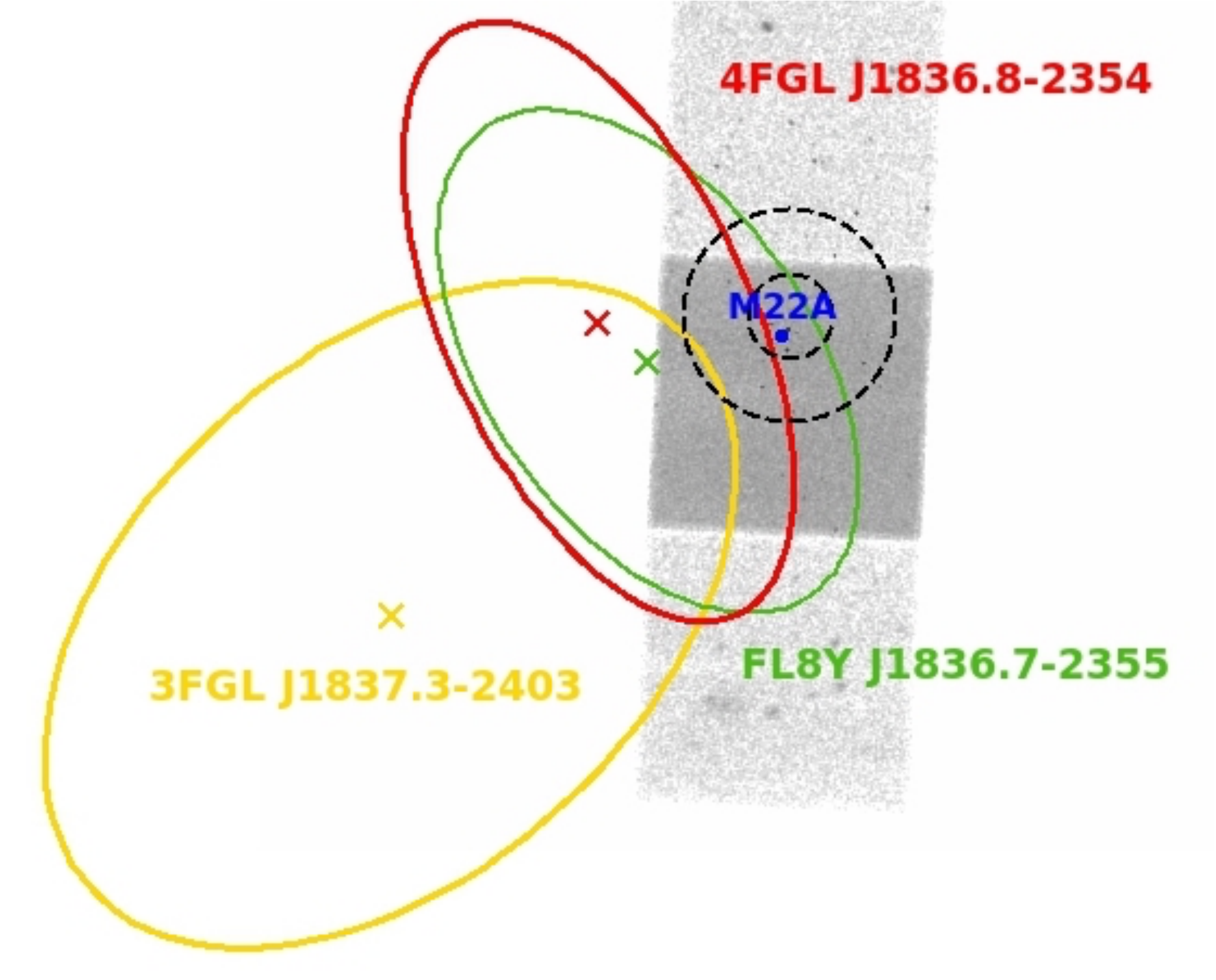}}
\caption{$\gamma$-ray sources and 95\% error ellipses from the 3FGL \protect\citep[in yellow,][]{Acero2015}, the preliminary FL8Y (in green) and the 4FGL \protect\citep[in red,][]{4FGL2019arXiv} catalogues of \fermi{}. The blue dot marks the radio position of the MSP M22A, while the black dashed circles the core radius (inner circle) and the half-mass radius (outer circle) of M22.}
\label{fig_fermi}
\end{figure}

Based on the $\gamma$-ray association to the GC M22 by \cite{zhou2015}, 
we checked whether this $\gamma$-ray source is compatible with the M22A position by using the latest \fermi{} catalogues.
We found in the 4-year catalogue \citep[3FGL,][]{Acero2015} that the source 3FGL J1837.3--2403 is positionally consistent with the emission detected by \cite{zhou2015}, but the MSP M22A is off from the 95\% error region (Fig.~\ref{fig_fermi}, yellow ellipse)\footnote{The other MSP identified by \cite{Lynch2011}, M22B, does not fall in the 95\% 3FGL J1837.3--2403 error ellipse either.}. The 95\% error ellipse touches the half-mass radius of the cluster, but does not cover the cluster core. 3FGL J1837.3--2403 showed a power law spectrum with photon index $2.40\pm0.14$ and a flux in the 0.1--100 GeV range of $(8.7\pm1.7)\times10^{-12}$ \ergcms{}, consistent with the best fit power law by \cite{zhou2015}. The corresponding $\gamma$-ray luminosity is $(10.6\pm2.1)\times10^{33}$ \ergsec{}, for a distance of 3.2 kpc. 
3FGL J1837.3-2403 appears rather stable, as also indicated by the low variability index of 43.73 reported in the catalogue \citep[see also][for details on variability]{Acero2015}.

From the inspection of the preliminary 8-yr \fermi{} source list (FL8Y), we found that 3FGL J1837.3--2403 is associated to FL8Y J1836.7--2355, whose detection is at 6.45$\sigma$ and at only 5.1$^\prime$ from the cluster centre. Though the 95\% error ellipse is smaller (Fig.~\ref{fig_fermi}, green ellipse), it includes both the radio positions of the two MSPs M22A and M22B and obviously precludes a clear association to any of them.

While the present work was under review stage, the final 8-year catalogue \citep[4FGL,][]{4FGL2019arXiv} was officially released. The new release refines the preliminary position of the FL8Y list. The closest source to M22 is 4FGL J1836.8--2354, detected at 8.2$\sigma$, at a distance of almost 6$^{\prime}$ from the cluster centre. Its 95\% error region barely touches the cluster core and does not encompasses M22A, neither at the radio or X-ray position, though it is very close (see Fig.~\ref{fig_fermi}, red ellipse)\footnote{The 95\% elliptic region of 4FGL J1836.8--2354 does not encompass M22B either.}. In the 4FGL catalogue the source spectrum is found to be best fit with a log-normal representation (LogParabola)\footnote{See \url{https://fermi.gsfc.nasa.gov/ssc/data/access/lat/8yr_catalog/}.}.  
The significance of the fit of a LogParabola over a power law is 4.2$\sigma$.
The energy flux in the 0.1--100 GeV range is $(4.1\pm0.9)\times10^{-12}$ \ergcms{} with a corresponding $\gamma$-ray luminosity of $(5.0\pm1.1)\times10^{33}$ \ergsec{}. 
The difference in flux between the 3FGL and 4FGL catalogues is consistent within 2$\sigma$.

\section{Discussion\label{discussion}}

In this work, we present a comprehensive study of the radio MSP M22A, located in the GC M22, from multiwavelength observations.
We search for X-ray emission from M22A, taking into account all the available X-ray observations within the last two decades.
Using the most recent \chandra\ observation of 2014, we detect an X-ray source whose 1$\sigma$ positional uncertainty encompasses the radio source M22A and therefore we ascribe it as the X-ray counterpart of the radio MSP. Thanks to its $\sim$\,85 ks of exposure time, the \chandra{} observation allows us to investigate the spectral shape and to determine the X-ray luminosity of the pulsar.
We do not detect any X-ray emission from M22A in either \xmm{} or \sw{}/XRT pointings; the \sw\ monitoring  campaign of the cluster, with one or two visits per month, shows that M22A remains likely around, or
below, the luminosity derived in the \chandra\ observations.\\

We studied the X-ray spectrum of M22A by using the data from the two \chandra\ observations. We considered two possible scenarios: a non-thermal emission, originating from an intrabinary shock produced between the powerful pulsar wind and that from the companion star \citep{romani16,Wadiasingh2017}, and a thermal emission, which could originate in the polar caps of the NS, where the infall of relativistic particles keeps heating the pulsar surface \citep{Gentile2014}. Both the emission mechanisms are discussed below. 

\begin{figure}
\resizebox{\hsize}{!}
{\includegraphics[width=.48\textwidth]{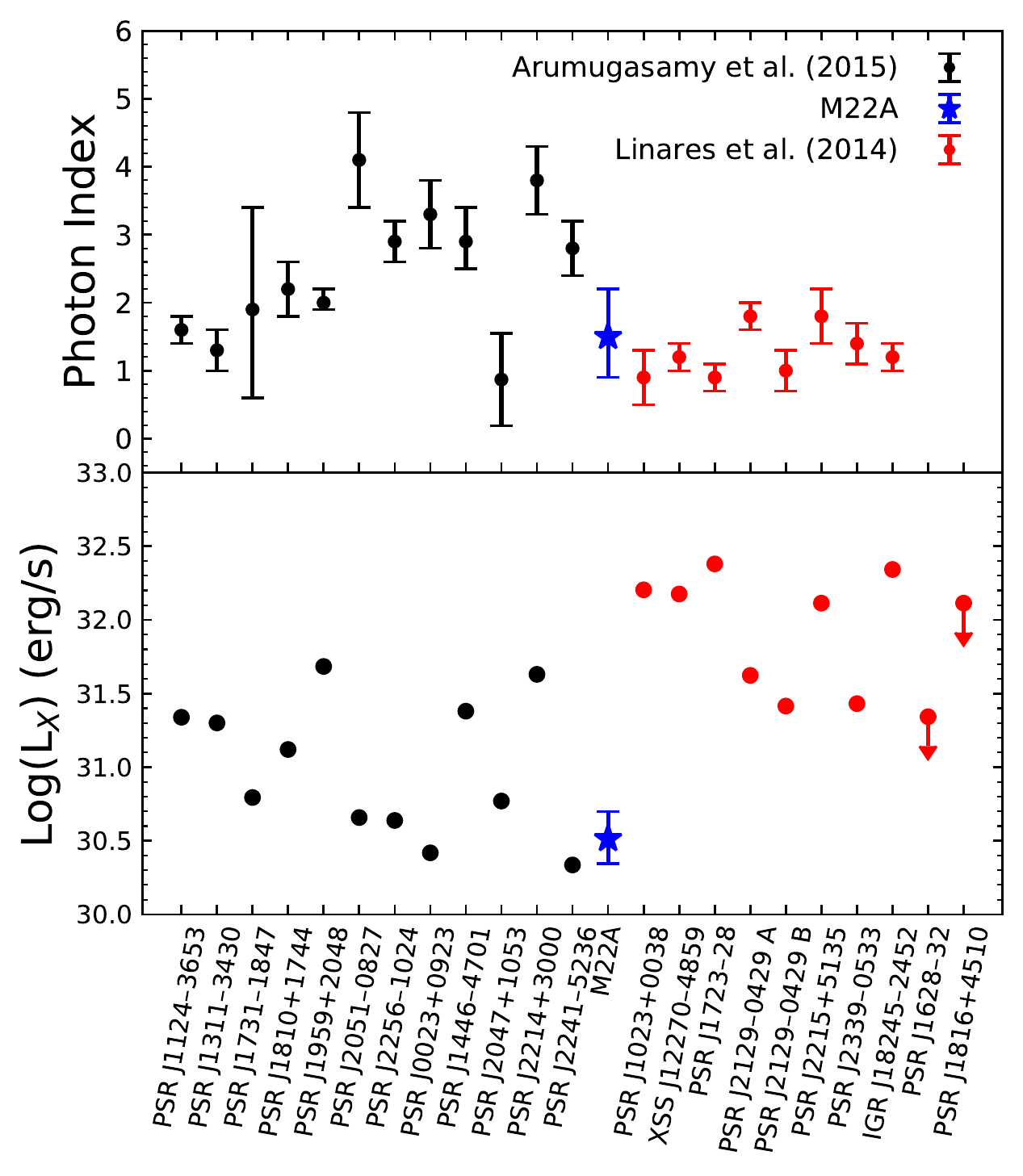}}
\caption{\bf Photon indices (\textit{top panel}) and X-ray luminosities in the energy range 0.5-10 keV (\textit{bottom panel}) of a sample of black widows (black) from \protect\cite{Arumugasamy2015}, redbacks (red) from \protect\cite{Linares2014} and our derived values for M22A (blue star).}
\label{comparison_gamma}
\end{figure}

The X-ray spectrum can be reasonably fitted with a relatively hard power-law ($\Gamma\sim$1.5) which could hint at a non-thermal origin and favours the intrabinary shock scenario. In fact, the X-ray emission from the shock is expected to be hard with a power law shape with index $1.1-1.2$ \citep{Becker1999,Zavlin2007}. The X-ray flux and spectrum  is also expected to be variable at the binary orbital period, as indeed found in most systems \citep{Bogdanov2005,Gentile2014,DeMartino2015,Roberts2015}. Unfortunately, due to the low statistics, we could not infer any orbital modulation. 
We compare the photon index of M22A with those presented by \cite{Arumugasamy2015} for a sample of black widow pulsars \citep[see also][]{Gentile2014} and those of \cite{Linares2014} for a sample of redbacks \citep[see also][]{Roberts2015,Strader2019}, as shown in Fig.~\ref{comparison_gamma} (top panel). Though the photon index of M22A is poorly constrained, it is consistent with similar hard values found in a number of black widows and in all redbacks.

Thermal emission is often observed from faint MSPs, where the total power generated is $log_{10}(L_{x})$=30-32 erg s$^{-1}$ \citep{Bogdanov2006,Forestell2014,Bhattacharya2017} and the magnetic field is low, typically $B\lesssim10^9$ G \citep{Zavlin1996,Heinke2006}. The intensity of the magnetic field at the surface of the NS, in the simple case of a magnetic dipole, is given by $B_{\text{surf}}=3.2\times10^{19}(P\dot{P})^{1/2}$ G \citep{Manchester1977}, where $P$ and $\dot{P}$ are respectively the spin period and the spin-down rate of the NS. From \cite{Lynch2011}, $P\simeq3.35$ ms and $\dot{P}\simeq5.36\times10^{-21}$ ss$^{-1}$, being $\dot{P}$ the intrinsic spin-down of the pulsar, disentangled from the effect due to the potential of the Galaxy and of the proper motion of the cluster \citep[formula 9]{Lynch2011}. Hence, $B_{\text{surf}}\sim1.4\times10^8$ G, implying that the contribution of a thermal emission cannot be excluded.

The X-ray spectrum, indeed, could be equally described by a black-body with temperature of $0.8\pm0.4$ keV. It is perfectly consistent with the temperatures of other samples of X-ray pulsars (see, for instance, \cite{Bogdanov2006} and \cite{Bhattacharya2017} for a spectral analysis of the MSPs of the GC 47 Tucanae).

\begin{figure}
\resizebox{\hsize}{!}
{\includegraphics[height=.3\textheight]{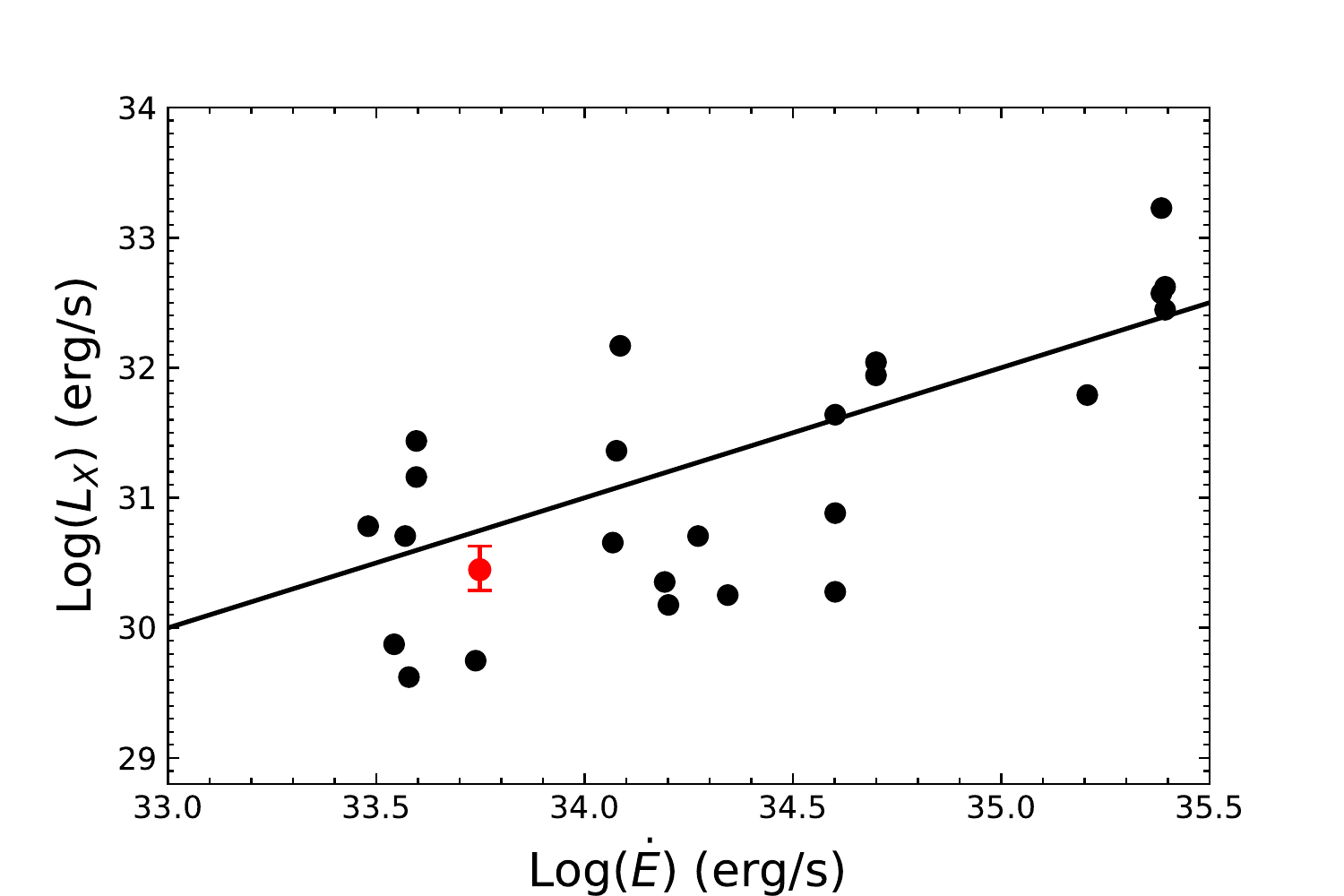}}
\caption{Luminosity versus spin-down energy loss rate $\dot{E}$ for 24 X-ray detected MSPs (black circles), from \protect\cite{Gentile2014}. The red dot stands for M22A. Luminosities are in the 0.3-8.0 keV range, while the line represents 0.1\% efficiency.}
\label{log_lx_lgamma}
\end{figure}

To argue more deeply about the thermal scenario, we can use the correlation between the X-ray luminosity and the rotational energy loss rate ($\dot{E}=4\pi^2I\dot{P}/P^3$), which is equal to $\sim5.6\times10^{33}$ \ergsec{} for M22A. We compare our result with a sample of 24 MSPs \citep{Gentile2014} in Fig.~\ref{log_lx_lgamma}. Under the hypothesis that the rotational energy loss rate is converted in X-ray thermal emission from the polar caps with an efficiency of 0.1\% \citep{Pavlov2007} (\textit{solid line} in Fig.~\ref{log_lx_lgamma}), the thermal conversion mechanism would seem to be plausible for M22A. However, we underline that the best-fit  value of the radius of the emitting polar cap, $R_{\text{eff}}=6.5^{+8}_{-4}$ m (Tab. \ref{tab_fit_results}), is unrealistically small.\\

We derive an X-ray luminosity of $(2-3)\times10^{30}$ \ergsec{}, for the black-body and the power law models, respectively, in the energy range 0.5--8 keV. These values are consistent with the ones typically found for GC X-ray sources ($L_X\sim10^{30}-10^{31}$ \ergsec{}) \citep{Bogdanov2006}. On the base of the X-ray luminosity, we  try to discriminate whether M22A is more likely a black widow or a redback. For this purpose, we made a comparison between the X-ray luminosities of the black widow pulsars from \cite{Arumugasamy2015} and of the redbacks from \cite{Linares2014}, as shown in the bottom panel of Fig~\ref{comparison_gamma} (for a wider sample consider also sources from \cite{Gentile2014}, \cite{Roberts2015} and \cite{Strader2019}). Black widows luminosities are in the range $\log_{10}(L_X)=30.2-31.3$ \ergsec{}, while redbacks luminosities seem to be sistematically higher, in the range $\log_{10}(L_X)=31.5-33.7$ \ergsec{}. With a value of $\log_{10}(L_X)=30.5$ \ergsec{}, in the range 0.5--10 keV, M22A is more consistent with black widows rather than with redbacks.

The persistent low X-ray flux does not favour accretion of matter from the companion star. The low companion mass and relatively large orbital period seem to indicate that mass accretion in this system is unlikely. The mass function of $2.6\times10^{-6}$ indicates a companion star of mass $M_2=0.017\,M_{\odot}$ for $i=90^\circ$ \citep{Lynch2011} and $M_2=0.22\,M_{\odot}$ for $i=5^\circ$. 
We exclude lower angles, being the probability of observing a binary system with an inclination $i\,<\,5^\circ$ equal to $1\,-\,\cos{(i)}\,\simeq\,0.4\%$ \citep{Lorimer2004Book}. 
Using $M_2=0.22\,M_{\odot}$ as an upper limit, we consider a Roche-lobe overflow as possible mechanism of mass transfer.  
In this case the secondary star radius $R_2$ must be at least of the same order of magnitude of its Roche lobe radius $R_L$, therefore it is sufficient to compare the two radii $R_2$ and $R_{L,2}$.
The size of the Roche lobes is
$R_{L_2}=0.49q^{2/3}/[0.6q^{2/3}+\ln(1+q^{1/3})]a$ \citep{eggleton83}, where $q$ is the ratio between $M_2$ and $M_1$, the mass of the primary star, and of the orbital separation $a$. We adopt a mass of 1.4 $M_\odot$ for the NS and the range 0.02--0.2 $M_\odot$ for the companion, according to the possible inclinations of the system. Using the third Kepler's law we derive an orbital separation $a$ in the range (1.14--1.16)$\times$10$^6$ km and, hence, $R_{L,2}=(1.2-2.7)\times10^5$ km (0.18--0.39$\,R_\odot$). On the other hand, an estimation of R$_2$ can be made according to the mass-radius relationships for low mass stars and sub-stellar objects by \cite{Chabrier2000Apj} (see their Table 5); 
for an ``old'' object, with an age of $\approx$10 Gyr and a mass between 0.05 and 0.1 $M_\odot$, the radius ranges between 0.08 and 0.12\,$R_\odot$, which is about the Jupiter radius. Since $R_2<R_{L_2}$, the accretion of matter onto the NS through Roche-lobe overflow is ruled out. 

However, it cannot be excluded that the companion star is out of thermal equilibrium and bloated with respect to its main sequence configuration \citep[see, e.g.,][]{King1988}. In this case, the companion star can be close to fill its Roche lobe and can transfer or loose mass (as it happens in red-backs and black widows) thanks also to the pulsar irradiation. In any case, we do not expect accretion in this phase of the system since the radiation pressure from the pulsar may be able to expel the mass transferred by the companion star out of the system  \citep[see, e.g.,][]{Burderi2001}.\\

Even in the case of a lack of detection of an optical counterpart, we can derive some constraints on the nature of the companion of M22A.
We compare the expected magnitudes for the case of maximum radii, i.e. Roche-lobe filling between 0.18 and 0.39\,$R_\odot$, adopting temperatures up to 3400 K. Here we note that no brown dwarf is expected to have temperatures above 3000 K and radius larger than 0.2 $R_\odot$ even at 0.1 Gyr \citep{Chabrier2000Apj}.  
The upper limits in the F606W and F814W filters derived
from \textit{HST}, once converted into the Johnson-Cousin system \citep{Sirianni2005} and
adopting an interstellar extinction E(B-V)=0.34 \citep{Alonsogarcia2012} and the distance of 3.2 kpc, give absolute magnitudes of 12.5 and 11.6 in the V and I bands, respectively. These values are well above the evolutionary sequences of brown dwarfs by more than 3 mag in V and 1 mag in I \citep{Chabrier2000Apj}. For $R_2$ between 0.18\,$R_\odot$ and 0.39\,$R_\odot$ and $T_{\text{eff}}=3400$ K, the expected magnitudes are V=13.3--10.8 mag and I=10.8--8.6 mag, respectively. On the other hand, the limits in the V and I bands would correspond, for a similar temperature, to a stellar radius of 0.23\,$R_\odot$ and 0.16\,$R_\odot$. In the case of Roche-lobe filling, i.e. $R_2$\,$=$\,$R_{L_2}$, adopting again 1.4\,$M_\odot$ for the NS, these radii would correspond to masses between 0.04 and 0.014
$M_\odot$, respectively. Releasing the Roche-lobe filling condition, the magnitude limits and thus the corresponding upper limits to the radii give a main sequence star of 0.2 $M_\odot$ and 0.1 $M_\odot$ respectively \citep{Baraffe2015}.
Therefore, although tentative, these estimate appears to rule out a companion with a mass above 0.1--0.2 $M_\odot$. According to the recent study of \cite{Strader2019}, redback
companions have median masses of 0.36$\pm$0.04 $M_\odot$, with a scatter of $\sigma$=0.15$\pm$0.04 $M_\odot$. Thus, our analysis may favour a black widow binary, in agreement with the interpretation of \cite{Lynch2011}.\\

Concerning the $\gamma$-ray emission, the new position and uncertainty in the 8-year catalogue seem to exclude the contribution of the two MSPs to the $\gamma$-ray emission of 4FGL J1836.8--2354, although the 95\% error ellipse is only slightly offset from the two radio sources.
The number of MSPs expected in the cluster can be estimated as $N_{\text{MSP}}=L_\gamma/\braket{\dot{E}}\braket{\eta_\gamma}$ \citep{Abdo2010}, where $L_\gamma$ is the $\gamma$-ray luminosity of the cluster, $\braket{\dot{E}}$ is the average power loss during the spin down of MSPs and $\braket{\eta_\gamma}$ is the average conversion efficiency of the spin down power into $\gamma$-ray radiation. Assuming  $\braket{\dot{E}}=(1.8\pm0.7)\times10^{34}$ \ergsec{}, $\braket{\eta_\gamma}=0.08$ \citep{Abdo2010} and $L_\gamma=5\times10^{33}$ \ergsec{}, we obtain $N_{\text{MSP}}\simeq4$, i.e. we expect that the $\gamma$-ray emission seen from \textit{Fermi} is the cumulative contribution of at least 4 MSPs. With only 2 radio MSPs detected in M22 so far, we are unable to assess their true contribution. The curved $\gamma$-ray spectrum, as reported in the 8-year \fermi{} catalogue, may be also compatible with an origin from pulsars \citep{4FGL2019arXiv}.
 
\section{Conclusions}
We have carried out a search for the X-ray, optical and $\gamma$-ray counterparts of the radio MSP M22A, detected by \cite{Lynch2011}. We find persistent X-ray emission in two \chandra{} observations, made in 2005 and 2014 respectively. The X-ray spectrum is well modeled either with a hard power law, with a photon index of $\sim$1.5, or with a black-body model with a temperature of $\sim$0.8 keV. However, the latter gives an unrealistic value of the effective polar cap radius, which makes the intrabinary shock scenario more likely than thermal emission from the NS surface.
No optical counterpart has been found and the inferred upper limits on the magnitudes allow us to derive an upper limit on the mass of the companion star of 0.2 $M_\odot$, typical for black widow systems. No $\gamma$-ray emission from M22 core is found in the latest \fermi{} catalogues. 

Further studies of this X-ray source can be made with new generation of satellites, like \textit{eRosita}, planned to flight in 2019, \textit{eXTP}, planned to flight earlier than 2025, or \textit{ATHENA}, whose launch is scheduled in 2030s, thus allowing more constraints on the nature of this system.

\section*{Acknowledgements}
We gratefully thank D. Nardiello and G.P. Piotto for providing positional accuracies
and background values of the \textit{HST} images within the HUGS project.
Authors acknowledge financial contribution from the agreement ASI-INAF n.2017-14-H.0. AD acknowledges support from the ASI Grant I/004/11/1. DDM acknowledges financial contribution from ``Towards the SKA and CTA era'' (P.I. M. Giroletti) and RI and TDS from the research grant ``iPeska'' (P.I. A. Possenti), both funded under the INAF national call Prin-SKA/CTA approved with the Presidential Decree 70/2016. RI and TDS aknowledge also financial contribution  from the HERMES Project, funded by the Italian Space Agency (ASI) Agreement n. 2016/13 U.O.

\bibliographystyle{mnras}
\bibliography{refs}

\begin{thebibliography}{}
\makeatletter
\relax
\def\mn@urlcharsother{\let\do\@makeother \do\$\do\&\do\#\do\^\do\_\do\%\do\~}
\def\mn@doi{\begingroup\mn@urlcharsother \@ifnextchar [ {\mn@doi@}
  {\mn@doi@[]}}
\def\mn@doi@[#1]#2{\def\@tempa{#1}\ifx\@tempa\@empty \href
  {http://dx.doi.org/#2} {doi:#2}\else \href {http://dx.doi.org/#2} {#1}\fi
  \endgroup}
\def\mn@eprint#1#2{\mn@eprint@#1:#2::\@nil}
\def\mn@eprint@arXiv#1{\href {http://arxiv.org/abs/#1} {{\tt arXiv:#1}}}
\def\mn@eprint@dblp#1{\href {http://dblp.uni-trier.de/rec/bibtex/#1.xml}
  {dblp:#1}}
\def\mn@eprint@#1:#2:#3:#4\@nil{\def\@tempa {#1}\def\@tempb {#2}\def\@tempc
  {#3}\ifx \@tempc \@empty \let \@tempc \@tempb \let \@tempb \@tempa \fi \ifx
  \@tempb \@empty \def\@tempb {arXiv}\fi \@ifundefined
  {mn@eprint@\@tempb}{\@tempb:\@tempc}{\expandafter \expandafter \csname
  mn@eprint@\@tempb\endcsname \expandafter{\@tempc}}}

\bibitem[\protect\citeauthoryear{{Abdo} et~al.,}{{Abdo}
  et~al.}{2010}]{Abdo2010}
{Abdo} A.~A.,  et~al., 2010, \mn@doi [\aap] {10.1051/0004-6361/201014458},
  \href {http://adsabs.harvard.edu/abs/2010A\%26A...524A..75A} {524, A75}

\bibitem[\protect\citeauthoryear{{Acero} et~al.,}{{Acero}
  et~al.}{2015}]{Acero2015}
{Acero} F.,  et~al., 2015, \mn@doi [\apjs] {10.1088/0067-0049/218/2/23}, \href
  {http://adsabs.harvard.edu/abs/2015ApJS..218...23A} {218, 23}

\bibitem[\protect\citeauthoryear{{Alonso-Garc{\'{\i}}a}, {Mateo}, {Sen},
  {Banerjee}, {Catelan}, {Minniti}  \& {von Braun}}{{Alonso-Garc{\'{\i}}a}
  et~al.}{2012}]{Alonsogarcia2012}
{Alonso-Garc{\'{\i}}a} J.,  {Mateo} M.,  {Sen} B.,  {Banerjee} M.,  {Catelan}
  M.,  {Minniti} D.,   {von Braun} K.,  2012, \mn@doi [\aj]
  {10.1088/0004-6256/143/3/70}, \href
  {http://adsabs.harvard.edu/abs/2012AJ....143...70A} {143, 70}

\bibitem[\protect\citeauthoryear{{Alpar}, {Cheng}, {Ruderman}  \&
  {Shaham}}{{Alpar} et~al.}{1982}]{Alpar1982}
{Alpar} M.~A.,  {Cheng} A.~F.,  {Ruderman} M.~A.,   {Shaham} J.,  1982, \mn@doi
  [\nat] {10.1038/300728a0}, \href
  {http://adsabs.harvard.edu/abs/1982Natur.300..728A} {300, 728}

\bibitem[\protect\citeauthoryear{{Archibald} et~al.,}{{Archibald}
  et~al.}{2009}]{Archibald2009}
{Archibald} A.~M.,  et~al., 2009, \mn@doi [Science] {10.1126/science.1172740},
  \href {http://adsabs.harvard.edu/abs/2009Sci...324.1411A} {324, 1411}

\bibitem[\protect\citeauthoryear{{Arumugasamy}, {Pavlov}  \&
  {Garmire}}{{Arumugasamy} et~al.}{2015}]{Arumugasamy2015}
{Arumugasamy} P.,  {Pavlov} G.~G.,   {Garmire} G.~P.,  2015, \mn@doi [\apj]
  {10.1088/0004-637X/814/2/90}, \href
  {http://adsabs.harvard.edu/abs/2015ApJ...814...90A} {814, 90}

\bibitem[\protect\citeauthoryear{{Baraffe}, {Homeier}, {Allard}  \&
  {Chabrier}}{{Baraffe} et~al.}{2015}]{Baraffe2015}
{Baraffe} I.,  {Homeier} D.,  {Allard} F.,   {Chabrier} G.,  2015, \mn@doi
  [\aap] {10.1051/0004-6361/201425481}, \href
  {http://adsabs.harvard.edu/abs/2015A%26A...577A..42B} {577, A42}

\bibitem[\protect\citeauthoryear{{Bassa} et~al.,}{{Bassa}
  et~al.}{2014}]{Bassa2014}
{Bassa} C.~G.,  et~al., 2014, \mn@doi [\mnras] {10.1093/mnras/stu708}, \href
  {http://adsabs.harvard.edu/abs/2014MNRAS.441.1825B} {441, 1825}

\bibitem[\protect\citeauthoryear{{Becker} \& {Tr{\"u}mper}}{{Becker} \&
  {Tr{\"u}mper}}{1999}]{Becker1999}
{Becker} W.,  {Tr{\"u}mper} J.,  1999, \aap, \href
  {http://adsabs.harvard.edu/abs/1999A%26A...341..803B} {341, 803}

\bibitem[\protect\citeauthoryear{{Bhattacharya} \& {van den
  Heuvel}}{{Bhattacharya} \& {van den
  Heuvel}}{1991}]{BhattacharyaVandenHeuvel1991}
{Bhattacharya} D.,  {van den Heuvel} E.~P.~J.,  1991, \mn@doi [\physrep]
  {10.1016/0370-1573(91)90064-S}, \href
  {http://adsabs.harvard.edu/abs/1991PhR...203....1B} {203, 1}

\bibitem[\protect\citeauthoryear{{Bhattacharya}, {Heinke}, {Chugunov},
  {Freire}, {Ridolfi}  \& {Bogdanov}}{{Bhattacharya}
  et~al.}{2017}]{Bhattacharya2017}
{Bhattacharya} S.,  {Heinke} C.~O.,  {Chugunov} A.~I.,  {Freire} P.~C.~C.,
  {Ridolfi} A.,   {Bogdanov} S.,  2017, \mn@doi [\mnras]
  {10.1093/mnras/stx2241}, \href
  {http://adsabs.harvard.edu/abs/2017MNRAS.472.3706B} {472, 3706}

\bibitem[\protect\citeauthoryear{{Bogdanov}, {Grindlay}  \& {van den
  Berg}}{{Bogdanov} et~al.}{2005}]{Bogdanov2005}
{Bogdanov} S.,  {Grindlay} J.~E.,   {van den Berg} M.,  2005, \mn@doi [\apj]
  {10.1086/432249}, \href {http://adsabs.harvard.edu/abs/2005ApJ...630.1029B}
  {630, 1029}

\bibitem[\protect\citeauthoryear{{Bogdanov}, {Grindlay}, {Heinke}, {Camilo},
  {Freire}  \& {Becker}}{{Bogdanov} et~al.}{2006}]{Bogdanov2006}
{Bogdanov} S.,  {Grindlay} J.~E.,  {Heinke} C.~O.,  {Camilo} F.,  {Freire}
  P.~C.~C.,   {Becker} W.,  2006, \mn@doi [\apj] {10.1086/505133}, \href
  {http://adsabs.harvard.edu/abs/2006ApJ...646.1104B} {646, 1104}

\bibitem[\protect\citeauthoryear{{Burderi} et~al.,}{{Burderi}
  et~al.}{2001}]{Burderi2001}
{Burderi} L.,  et~al., 2001, \mn@doi [\apjl] {10.1086/324220}, \href
  {http://adsabs.harvard.edu/abs/2001ApJ...560L..71B} {560, L71}

\bibitem[\protect\citeauthoryear{{Burrows}, {Hill}, {Nousek}, {Kennea}, {Wells}
   \& {Osborne}}{{Burrows} et~al.}{2005}]{burrows05}
{Burrows} D.~N.,  {Hill} J.~E.,  {Nousek} J.~A.,  {Kennea} J.~A.,  {Wells} A.,
   {Osborne} 2005, \mn@doi [Space Science Reviews] {10.1007/s11214-005-5097-2},
  \href {http://adsabs.harvard.edu/abs/2005SSRv..120..165B} {120, 165}

\bibitem[\protect\citeauthoryear{{Caraveo}}{{Caraveo}}{2014}]{Caraveo2014}
{Caraveo} P.~A.,  2014, \mn@doi [\araa] {10.1146/annurev-astro-081913-035948},
  \href {http://adsabs.harvard.edu/abs/2014ARA%26A..52..211C} {52, 211}

\bibitem[\protect\citeauthoryear{{Cash}}{{Cash}}{1979}]{Cash1979}
{Cash} W.,  1979, \mn@doi [\apj] {10.1086/156922}, \href
  {http://adsabs.harvard.edu/abs/1979ApJ...228..939C} {228, 939}

\bibitem[\protect\citeauthoryear{{Chabrier}, {Baraffe}, {Allard}  \&
  {Hauschildt}}{{Chabrier} et~al.}{2000}]{Chabrier2000Apj}
{Chabrier} G.,  {Baraffe} I.,  {Allard} F.,   {Hauschildt} P.,  2000, \mn@doi
  [\apj] {10.1086/309513}, \href
  {http://adsabs.harvard.edu/abs/2000ApJ...542..464C} {542, 464}

\bibitem[\protect\citeauthoryear{{Chen}}{{Chen}}{1991}]{Chen1991}
{Chen} K.,  1991, \mn@doi [\nat] {10.1038/352695a0}, \href
  {http://adsabs.harvard.edu/abs/1991Natur.352..695C} {352, 695}

\bibitem[\protect\citeauthoryear{{Cheng}, {Li}, {Xu}  \& {Li}}{{Cheng}
  et~al.}{2018}]{Cheng2018}
{Cheng} Z.,  {Li} Z.,  {Xu} X.,   {Li} X.,  2018, \mn@doi [APJ]
  {10.3847/1538-4357/aaba16}, \href
  {http://adsabs.harvard.edu/abs/2018ApJ...858...33C} {858, 33}

\bibitem[\protect\citeauthoryear{{Eggleton}}{{Eggleton}}{1983}]{eggleton83}
{Eggleton} P.~P.,  1983, \mn@doi [\apj] {10.1086/160960}, \href
  {http://adsabs.harvard.edu/abs/1983ApJ...268..368E} {268, 368}

\bibitem[\protect\citeauthoryear{{Fabian}, {Pringle}  \& {Rees}}{{Fabian}
  et~al.}{1975}]{Fabian1975}
{Fabian} A.~C.,  {Pringle} J.~E.,   {Rees} M.~J.,  1975, \mn@doi [\mnras]
  {10.1093/mnras/172.1.15P}, \href
  {http://adsabs.harvard.edu/abs/1975MNRAS.172P..15F} {172, 15p}

\bibitem[\protect\citeauthoryear{{Forbes} \& {Bridges}}{{Forbes} \&
  {Bridges}}{2010}]{forbes10}
{Forbes} D.~A.,  {Bridges} T.,  2010, \mn@doi [MNRAS]
  {10.1111/j.1365-2966.2010.16373.x}, \href
  {http://adsabs.harvard.edu/abs/2010MNRAS.404.1203F} {404, 1203}

\bibitem[\protect\citeauthoryear{{Forestell}, {Heinke}, {Cohn}, {Lugger},
  {Sivakoff}, {Bogdanov}, {Cool}  \& {Anderson}}{{Forestell}
  et~al.}{2014}]{Forestell2014}
{Forestell} L.~M.,  {Heinke} C.~O.,  {Cohn} H.~N.,  {Lugger} P.~M.,  {Sivakoff}
  G.~R.,  {Bogdanov} S.,  {Cool} A.~M.,   {Anderson} J.,  2014, \mn@doi
  [\mnras] {10.1093/mnras/stu559}, \href
  {http://adsabs.harvard.edu/abs/2014MNRAS.441..757F} {441, 757}

\bibitem[\protect\citeauthoryear{{Gehrels}, {Chincarini}, {Giommi}, {Mason},
  {Nousek}, {Wells}  \& {White}}{{Gehrels} et~al.}{2004}]{gehrels04}
{Gehrels} N.,  {Chincarini} G.,  {Giommi} P.,  {Mason} K.~O.,  {Nousek} J.~A.,
  {Wells} A.~A.,   {White} N.~E.,  2004, \mn@doi [\apj] {10.1086/422091}, \href
  {http://adsabs.harvard.edu/abs/2004ApJ...611.1005G} {611, 1005}

\bibitem[\protect\citeauthoryear{{Geller} et~al.,}{{Geller}
  et~al.}{2017}]{Geller2017}
{Geller} A.~M.,  et~al., 2017, \mn@doi [\apj] {10.3847/1538-4357/aa6af3}, \href
  {http://adsabs.harvard.edu/abs/2017ApJ...840...66G} {840, 66}

\bibitem[\protect\citeauthoryear{{Gentile} et~al.,}{{Gentile}
  et~al.}{2014}]{Gentile2014}
{Gentile} P.~A.,  et~al., 2014, \mn@doi [\apj] {10.1088/0004-637X/783/2/69},
  \href {http://adsabs.harvard.edu/abs/2014ApJ...783...69G} {783, 69}

\bibitem[\protect\citeauthoryear{{Harding}, {Usov}  \& {Muslimov}}{{Harding}
  et~al.}{2005}]{Harding2005}
{Harding} A.~K.,  {Usov} V.~V.,   {Muslimov} A.~G.,  2005, \mn@doi [\apj]
  {10.1086/427840}, \href {http://adsabs.harvard.edu/abs/2005ApJ...622..531H}
  {622, 531}

\bibitem[\protect\citeauthoryear{{Harris}}{{Harris}}{1996}]{Harris1996}
{Harris} W.~E.,  1996, \mn@doi [AJ] {10.1086/118116}, \href
  {http://adsabs.harvard.edu/abs/1996AJ....112.1487H} {112, 1487}

\bibitem[\protect\citeauthoryear{{Heinke}}{{Heinke}}{2010}]{Heinke2010}
{Heinke} C.~O.,  2010, in {Kalogera} V.,  {van der Sluys} M.,  eds,  American
  Institute of Physics Conference Series Vol. 1314, American Institute of
  Physics Conference Series. pp 135--142 (\mn@eprint {arXiv} {1101.5356}),
  \mn@doi{10.1063/1.3536355}

\bibitem[\protect\citeauthoryear{{Heinke}, {Rybicki}, {Narayan}  \&
  {Grindlay}}{{Heinke} et~al.}{2006}]{Heinke2006}
{Heinke} C.~O.,  {Rybicki} G.~B.,  {Narayan} R.,   {Grindlay} J.~E.,  2006,
  \mn@doi [\apj] {10.1086/503701}, \href
  {http://adsabs.harvard.edu/abs/2006ApJ...644.1090H} {644, 1090}

\bibitem[\protect\citeauthoryear{{Hertz} \& {Grindlay}}{{Hertz} \&
  {Grindlay}}{1983}]{Hertz1983}
{Hertz} P.,  {Grindlay} J.~E.,  1983, \mn@doi [\apj] {10.1086/161516}, \href
  {http://adsabs.harvard.edu/abs/1983ApJ...275..105H} {275, 105}

\bibitem[\protect\citeauthoryear{{Hills}}{{Hills}}{1976}]{Hills1976}
{Hills} J.~G.,  1976, \mn@doi [\mnras] {10.1093/mnras/175.1.1P}, \href
  {http://adsabs.harvard.edu/abs/1976MNRAS.175P...1H} {175, 1P}

\bibitem[\protect\citeauthoryear{{Johnston}, {Verbunt}  \&
  {Hasinger}}{{Johnston} et~al.}{1994}]{Johnston1994}
{Johnston} H.~M.,  {Verbunt} F.,   {Hasinger} G.,  1994, \aap, \href
  {http://adsabs.harvard.edu/abs/1994A%26A...289..763J} {289, 763}

\bibitem[\protect\citeauthoryear{{King}}{{King}}{1988}]{King1988}
{King} A.~R.,  1988, \qjras, \href
  {http://adsabs.harvard.edu/abs/1988QJRAS..29....1K} {29, 1}

\bibitem[\protect\citeauthoryear{{Linares}}{{Linares}}{2014}]{Linares2014}
{Linares} M.,  2014, \mn@doi [\apj] {10.1088/0004-637X/795/1/72}, \href
  {http://adsabs.harvard.edu/abs/2014ApJ...795...72L} {795, 72}

\bibitem[\protect\citeauthoryear{{Lorimer} \& {Kramer}}{{Lorimer} \&
  {Kramer}}{2004}]{Lorimer2004Book}
{Lorimer} D.~R.,  {Kramer} M.,  2004, {Handbook of Pulsar Astronomy}

\bibitem[\protect\citeauthoryear{{Lynch}, {Ransom}, {Freire}  \&
  {Stairs}}{{Lynch} et~al.}{2011}]{Lynch2011}
{Lynch} R.~S.,  {Ransom} S.~M.,  {Freire} P.~C.~C.,   {Stairs} I.~H.,  2011,
  \mn@doi [Astrophysical Journal] {10.1088/0004-637X/734/2/89}, \href
  {http://adsabs.harvard.edu/abs/2011ApJ...734...89L} {734, 89}

\bibitem[\protect\citeauthoryear{{Manchester} \& {Taylor}}{{Manchester} \&
  {Taylor}}{1977}]{Manchester1977}
{Manchester} R.~N.,  {Taylor} J.~H.,  1977, {Pulsars}

\bibitem[\protect\citeauthoryear{{Nardiello} et~al.,}{{Nardiello}
  et~al.}{2018}]{Nardiello2018}
{Nardiello} D.,  et~al., 2018, \mn@doi [\mnras] {10.1093/mnras/sty2515}, \href
  {http://adsabs.harvard.edu/abs/2018MNRAS.481.3382N} {481, 3382}

\bibitem[\protect\citeauthoryear{{Papitto} et~al.,}{{Papitto}
  et~al.}{2013}]{Papitto2013Nature}
{Papitto} A.,  et~al., 2013, \mn@doi [\nat] {10.1038/nature12470}, \href
  {http://adsabs.harvard.edu/abs/2013Natur.501..517P} {501, 517}

\bibitem[\protect\citeauthoryear{{Pavlov}, {Kargaltsev}, {Garmire}  \&
  {Wolszczan}}{{Pavlov} et~al.}{2007}]{Pavlov2007}
{Pavlov} G.~G.,  {Kargaltsev} O.,  {Garmire} G.~P.,   {Wolszczan} A.,  2007,
  \mn@doi [\apj] {10.1086/518926}, \href
  {http://adsabs.harvard.edu/abs/2007ApJ...664.1072P} {664, 1072}

\bibitem[\protect\citeauthoryear{{Piotto} et~al.,}{{Piotto}
  et~al.}{2015}]{Piotto2015}
{Piotto} G.,  et~al., 2015, \mn@doi [\aj] {10.1088/0004-6256/149/3/91}, \href
  {http://adsabs.harvard.edu/abs/2015AJ....149...91P} {149, 91}

\bibitem[\protect\citeauthoryear{{Roberts}, {McLaughlin}, {Gentile}, {Ray},
  {Ransom}  \& {Hessels}}{{Roberts} et~al.}{2015}]{Roberts2015}
{Roberts} M.~S.~E.,  {McLaughlin} M.~A.,  {Gentile} P.~A.,  {Ray} P.~S.,
  {Ransom} S.~M.,   {Hessels} J.~W.~T.,  2015, preprint, \href
  {http://adsabs.harvard.edu/abs/2015arXiv150207208R} {} (\mn@eprint {arXiv}
  {1502.07208})

\bibitem[\protect\citeauthoryear{{Roberts} et~al.,}{{Roberts}
  et~al.}{2018}]{roberts2018}
{Roberts} M.~S.~E.,  et~al., 2018, in {Weltevrede} P.,  {Perera} B.~B.~P.,
  {Preston} L.~L.,   {Sanidas} S.,  eds,  IAU Symposium Vol. 337, Pulsar
  Astrophysics the Next Fifty Years. pp 43--46 (\mn@eprint {arXiv}
  {1801.09903}), \mn@doi{10.1017/S1743921318000480}

\bibitem[\protect\citeauthoryear{{Romani} \& {Sanchez}}{{Romani} \&
  {Sanchez}}{2016}]{romani16}
{Romani} R.~W.,  {Sanchez} N.,  2016, \mn@doi [\apj]
  {10.3847/0004-637X/828/1/7}, \href
  {http://adsabs.harvard.edu/abs/2016ApJ...828....7R} {828, 7}

\bibitem[\protect\citeauthoryear{{Shishkovsky} et~al.,}{{Shishkovsky}
  et~al.}{2018}]{Shishkovsky2018}
{Shishkovsky} L.,  et~al., 2018, \mn@doi [\apj] {10.3847/1538-4357/aaadb1},
  \href {http://adsabs.harvard.edu/abs/2018ApJ...855...55S} {855, 55}

\bibitem[\protect\citeauthoryear{{Simioni} et~al.,}{{Simioni}
  et~al.}{2018}]{Simioni2018}
{Simioni} M.,  et~al., 2018, \mn@doi [\mnras] {10.1093/mnras/sty177}, \href
  {http://adsabs.harvard.edu/abs/2018MNRAS.476..271S} {476, 271}

\bibitem[\protect\citeauthoryear{{Sirianni} et~al.,}{{Sirianni}
  et~al.}{2005}]{Sirianni2005}
{Sirianni} M.,  et~al., 2005, \mn@doi [\pasp] {10.1086/444553}, \href
  {http://adsabs.harvard.edu/abs/2005PASP..117.1049S} {117, 1049}

\bibitem[\protect\citeauthoryear{{Stappers} et~al.,}{{Stappers}
  et~al.}{2014}]{Stappers2014}
{Stappers} B.~W.,  et~al., 2014, \mn@doi [\apj] {10.1088/0004-637X/790/1/39},
  \href {http://adsabs.harvard.edu/abs/2014ApJ...790...39S} {790, 39}

\bibitem[\protect\citeauthoryear{{Strader} et~al.,}{{Strader}
  et~al.}{2019}]{Strader2019}
{Strader} J.,  et~al., 2019, \mn@doi [\apj] {10.3847/1538-4357/aafbaa}, \href
  {http://adsabs.harvard.edu/abs/2019ApJ...872...42S} {872, 42}

\bibitem[\protect\citeauthoryear{{Sutantyo}}{{Sutantyo}}{1975}]{Sutantyo1975}
{Sutantyo} W.,  1975, \aap, \href
  {http://adsabs.harvard.edu/abs/1975A\%26A....44..227S} {44, 227}

\bibitem[\protect\citeauthoryear{{The Fermi-LAT collaboration}}{{The Fermi-LAT
  collaboration}}{2019}]{4FGL2019arXiv}
{The Fermi-LAT collaboration} 2019, arXiv e-prints, \href
  {https://ui.adsabs.harvard.edu/\#abs/2019arXiv190210045C} {p.
  arXiv:1902.10045}

\bibitem[\protect\citeauthoryear{{Verner}, {Ferland}, {Korista}  \&
  {Yakovlev}}{{Verner} et~al.}{1996}]{Verner}
{Verner} D.~A.,  {Ferland} G.~J.,  {Korista} K.~T.,   {Yakovlev} D.~G.,  1996,
  \mn@doi [Astrophysical Journal] {10.1086/177435}, \href
  {http://adsabs.harvard.edu/abs/1996ApJ...465..487V} {465, 487}

\bibitem[\protect\citeauthoryear{{Wadiasingh}, {Harding}, {Venter},
  {B{\"o}ttcher}  \& {Baring}}{{Wadiasingh} et~al.}{2017}]{Wadiasingh2017}
{Wadiasingh} Z.,  {Harding} A.~K.,  {Venter} C.,  {B{\"o}ttcher} M.,   {Baring}
  M.~G.,  2017, \mn@doi [\apj] {10.3847/1538-4357/aa69bf}, \href
  {http://adsabs.harvard.edu/abs/2017ApJ...839...80W} {839, 80}

\bibitem[\protect\citeauthoryear{{Webb} \& {Servillat}}{{Webb} \&
  {Servillat}}{2013}]{Webb2013}
{Webb} N.~A.,  {Servillat} M.,  2013, \mn@doi [A\&A]
  {10.1051/0004-6361/201117229}, \href
  {http://adsabs.harvard.edu/abs/2013A%26A...551A..60W} {551, A60}

\bibitem[\protect\citeauthoryear{{Webb}, {Gendre}  \& {Barret}}{{Webb}
  et~al.}{2002}]{Webb2002}
{Webb} N.~A.,  {Gendre} B.,   {Barret} D.,  2002, \mn@doi [A\&A]
  {10.1051/0004-6361:20011501}, \href
  {http://adsabs.harvard.edu/abs/2002A%26A...381..481W} {381, 481}

\bibitem[\protect\citeauthoryear{{Webb}, {Serre}, {Gendre}, {Barret}, {Lasota}
  \& {Rizzi}}{{Webb} et~al.}{2004}]{Webb2004}
{Webb} N.~A.,  {Serre} D.,  {Gendre} B.,  {Barret} D.,  {Lasota} J.-P.,
  {Rizzi} L.,  2004, \mn@doi [A\&A] {10.1051/0004-6361:20040399}, \href
  {http://adsabs.harvard.edu/abs/2004A%26A...424..133W} {424, 133}

\bibitem[\protect\citeauthoryear{{Wilms}, {Allen}  \& {McCray}}{{Wilms}
  et~al.}{2000}]{Wilms}
{Wilms} J.,  {Allen} A.,   {McCray} R.,  2000, \mn@doi [Astrophysical Journal]
  {10.1086/317016}, \href {http://adsabs.harvard.edu/abs/2000ApJ...542..914W}
  {542, 914}

\bibitem[\protect\citeauthoryear{{Zavlin}}{{Zavlin}}{2007}]{Zavlin2007}
{Zavlin} V.~E.,  2007, \mn@doi [\apss] {10.1007/s10509-007-9297-y}, \href
  {http://adsabs.harvard.edu/abs/2007Ap%26SS.308..297Z} {308, 297}

\bibitem[\protect\citeauthoryear{{Zavlin}, {Pavlov}  \& {Shibanov}}{{Zavlin}
  et~al.}{1996}]{Zavlin1996}
{Zavlin} V.~E.,  {Pavlov} G.~G.,   {Shibanov} Y.~A.,  1996, \aap, \href
  {http://adsabs.harvard.edu/abs/1996A%26A...315..141Z} {315, 141}

\bibitem[\protect\citeauthoryear{{Zhou}, {Zhang}, {Huang}, {Li}, {Liang}, {Fu},
  {Yan}  \& {Liu}}{{Zhou} et~al.}{2015}]{zhou2015}
{Zhou} J.~N.,  {Zhang} P.~F.,  {Huang} X.~Y.,  {Li} X.,  {Liang} Y.~F.,  {Fu}
  L.,  {Yan} J.~Z.,   {Liu} Q.~Z.,  2015, \mn@doi [\mnras]
  {10.1093/mnras/stv185}, \href
  {http://adsabs.harvard.edu/abs/2015MNRAS.448.3215Z} {448, 3215}

\bibitem[\protect\citeauthoryear{{de Martino} et~al.,}{{de Martino}
  et~al.}{2015}]{DeMartino2015}
{de Martino} D.,  et~al., 2015, \mn@doi [\mnras] {10.1093/mnras/stv2109}, \href
  {http://adsabs.harvard.edu/abs/2015MNRAS.454.2190D} {454, 2190}

\makeatother
\end{thebibliography}

\label{lastpage}
\end{document}